\documentclass[prb,altaffillsymbol,superscriptaddress, 
citeautoscript,amsfonts,amssymb,amsmath,twocolumn,floatfix,altaffillsymbol]
{revtex4}

\usepackage{graphicx}
\begin{document}

\date{\today}

\title{Electronic properties of the novel 4$d$ metallic oxide SrRhO$_3$}

\author{K. Yamaura}
\email[E-mail at:]{YAMAURA.Kazunari@nims.go.jp}
\homepage[Fax.:]{+81-298-58-5650}
\affiliation{Superconducting Materials Center, National Institute for Materials
Science, 1-1 Namiki, Tsukuba, Ibaraki 305-0044, Japan}

\author{Q. Huang}
\affiliation{NIST Center for Neutron Research, National Institute of Standards
and Technology, Gaithersburg, Maryland 20899}
\affiliation{Department of Materials and Nuclear Engineering, University of 
Maryland, College Park, Maryland 20742}

\author{D.P. Young}
\affiliation{Department of Physics and Astronomy, Louisiana State University,
Baton Rouge, LA 70803}

\author{M. Arai}
\affiliation{Computational Materials Science Center, National Institute
for Materials Science, 1-1 Namiki, Tsukuba, Ibaraki 305-0044, Japan}

\author{E. Takayama-Muromachi}
\affiliation{Superconducting Materials Center, National Institute for Materials
Science, 1-1 Namiki, Tsukuba, Ibaraki 305-0044, Japan}

\begin{abstract}
The novel 4$d$ perovskite compound SrRhO$_3$ was investigated by isovalent 
doping studies. 
The solubility limits of Ca and Ba onto Sr-site were below 80\% and 
20\%, respectively.
Although SrRhO$_3$ was chemically compressed, approximately 5.7\% by the Ca 
doping, no significant influence was observed on the magnetic and electrical
properties.
\end{abstract}

\maketitle


Novel 4$d$ electronic compounds in the rhodium-oxide system with perovskite-
and Ruddlesden-Popper-type structures were found recently, followed by 
intensive experimental investigations \cite{PRB01KY,PRB02KY,MRS02KY}.
An essential chemical reaction in synthesis of the compounds was provoked
by a high-pressure and high-temperature heating (6 GPa and 1500 $^\circ$C) in
our originally developed apparatus \cite{PRB01KY}.
The structure characteristics of the compounds were studied by means of powder 
neutron and x-ray diffraction \cite{PRB01KY,PRB02KY}; The structural
data clearly indicate them to be isostructural to the analogous ruthenium 
oxides with approximately the same degree of local structural distortions. 
The perovskite SrRhO$_3$ is metallic with enhanced paramagnetism as is
the analogous ruthenium oxide CaRuO$_3$ \cite{PRL99KY}.
A comprehensive picture, however, for the magnetic and transport properties
has not been fully established yet.
The quadratic temperature dependence, for example, of the magnetic 
susceptibility data of SrRhO$_3$ is unexpected, and it does not follow, even
qualitatively, the models of conventional paramagnetism or 
self-consistent-renormalization \cite{PRB01KY}.

In this short paper, we report the data of isovalent substitution studies on the
perovskite SrRhO$_3$.
The Ca substitution was achieved onto the Sr site up to approximately 80\%, 
resulting in 5.7\% compression in unit-cell volume.


Variable composition precursors Sr$_{1-x}$Ca$_x$RhO$_z$ ($x =$ 0 to 1 in 
0.2 steps) were prepared from SrCO$_3$ (99.9 \%), CaCO$_3$ (99.9 \%) and Rh 
(99.9 \%) powders. 
Mixtures were heated at 1200 $^\circ$C for 48 hrs in oxygen after a couple of
pre-heatings.
Each of those ($\sim$0.3 g) was then mixed with KClO$_4$ (8 wt.\%), and placed
into Pt capsules.
The capsules were compressed at 6 GPa and heated at 1500 $^\circ$C for 1 hr,
followed by quenching to room temperature at the elevated pressure.
Quality of the final products was examined by powder x-ray diffraction in a 
regular manner.
The magnetic susceptibility of the selected samples was measured in a 
commercial apparatus (Quantum Design, MPMS-XL).
The electrical resistivity was measured by a conventional dc-four-terminal 
technique.

\begin{figure}[btp]
\begin{center}
\leavevmode
\includegraphics[width=0.8\linewidth]{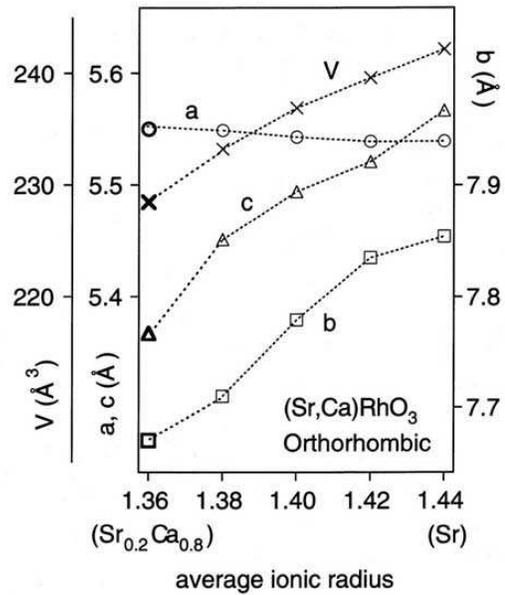}
\caption{Lattice parameters of the orthorhombic perovskite (Sr,Ca)RhO$_3$. Set
of the fat makers ($x=$ 0.8) is slightly out of solid-solution range.}
\label{fig.1}
\end{center}
\end{figure}

It appeared that the Ca doped samples ($x=$0.2--0.6) were of high-quality as
well as pure SrRhO$_3$ \cite{PRB01KY}.
At $x=$0.8, a small fraction of an unknown phase was detected, indicating a 
limit on the Ca solubility.
At the Ca-end ($x=$ 1.0), the sample consisted of multiple phases, which were
unidentified.
The various lattice parameters and the unit-cell volumes measured in the x-ray
study are arranged in Fig.\ref{fig.1}.
They decrease smoothly with increasing Ca concentration, consistent with Ca 
having a smaller ionic radius than Sr. 
The perovskite SrRhO$_3$ was chemically compressed $\sim5.7\%$ by the Ca 
substitution.

The temperature dependence of the electrical resistivity of the sample pellets
were measured between 2 K and 380 K.
Metallic behavior was observed for the samples between $x =$ 0 and 0.8 in the
temperature range.
Although the data were probably influenced somewhat by polycrystalline nature
of the samples, the essential electrical characteristics should be metallic 
over the whole solid-solution.

\begin{figure}[btp]
\begin{center}
\leavevmode
\includegraphics[width=0.8\linewidth]{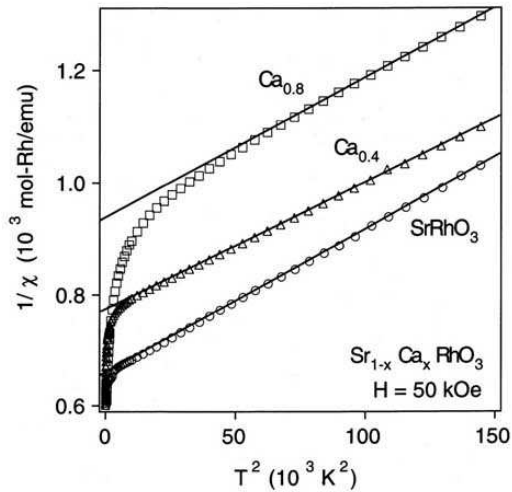}
\caption{The inverse magnetic susceptibility vs temperature squared at 50 kOe
for the polycrystalline samples of (Sr,Ca)RhO$_3$. The plots show a notably 
linear dependence as the solid lines indicate. Contribution from the sample 
holder was negligible. The Curie-Weiss law ($1/\chi \sim T$) was unable to fit
the magnetic data.}
\label{fig.2}
\end{center}
\end{figure}

The Ca concentration dependence of the magnetic properties was studied at 50
kOe between 2 K and 380 K. 
The magnetic susceptibility data could not be well fit to a typical 
Curie-Weiss type expression ($1/\chi \sim T$).
However, a linear region in the data is observed when plotted as $1/\chi$ vs
$T^2$ as shown in Fig.\ref{fig.2}.
To a first approximation, there is no change in slope of the linear part
of the data with increasing Ca concentration, rather generally a rigid shift
to higher values of $1/\chi$.
Neither antiferromagnetic nor ferromagnetic order was observed, and therefore,
the data do not provide sufficient evidence to determine the dominant 
influence on the rather unusual magnetic character ($1/\chi \sim T^2$) in the
metallic state.

The Ca-doping shifts the system away from a long-range magnetically ordered 
state, as the intersection between the horizontal axis and the extrapolated 
linear fit (Fig.\ref{fig.2}) moves away from the origin with increasing Ca 
concentration.
Long-range order is expected to appear when the point intersects the origin,
as found in the solid solution of the Ru analogue, (Ca,Sr)RuO$_3$ \cite
{PRL99KY}.
We were then motivated to try Ba-doping in the perovskite, essentially a study
in negative compression (Ba has a lager ionic size than Sr).
The amount of Ba substituted was, however, too insignificant to test the 
expectation.
The orthorhombic structure quickly transformed to a hexagonal type with 
increasing Ba concentration \cite{JSSC81BLC}; the Ba-solubility limit was less
than 20 \% at the synthesis conditions.

In summary, we reviewed investigations of the isovalent doping studies on 
SrRhO$_3$.
The data indicate no remarkable change in the magnetic and electrical 
properties of SrRhO$_3$, either qualitatively or quantitatively; the rather 
unusual magnetic character, quadratic temperature dependence of the magnetic
susceptibility ($1/\chi \sim T^2$) was robust against the 5.7\% chemical 
compression.
Hence, the open question still remains as to what mechanism is responsible for
the magnetic characteristics.
Further investigations, including testing aliovalent doping effects on 
SrRhO$_3$, would be of interest.

We wish to thank Dr. M. Akaishi (NIMS) and Dr. S. Yamaoka (NIMS) for their 
advice on the high-pressure experiments.

%
%


\begin{thebibliography}{9}
\bibitem{PRB01KY}
K. Yamaura, E. Takayama-Muromachi, Phys. Rev. B{\bf 64} (2001) 224424.
\bibitem{PRB02KY}
K. Yamaura, Q. Huang, D.P. Young, Y. Noguchi, E. Takayama-Muromachi, Phys. 
Rev. B (in press, cond-mat/0208467).
\bibitem{MRS02KY}
K. Yamaura, D. P. Young, and E. Takayama-Muromachi, in the 2002 MRS Spring 
Meeting,  San Francisco, California (in press).
\bibitem{PRL99KY}
K. Yoshimura, T. Imai, T. Kiyama, K.R. Thurber, A.W. Hunt, K. Kosuge, Phys. 
Rev. Lett. {\bf 83} (1999) 4397. 
\bibitem{JSSC81BLC}
B.L. Chamberland, J.B. Anderson, J. Solid State Chem. {\bf 39} (1981) 114.
\end{thebibliography}
\end{document}